\documentclass{sig-alternate}

\usepackage{amsmath}
\usepackage{mathtools}
\usepackage{color}
\usepackage{subcaption}
\usepackage{graphicx}
\usepackage{url}

\usepackage[numbers,sort]{natbib}

\usepackage{array}
\newcolumntype{L}[1]{>{\raggedright\let\newline\\\arraybackslash\hspace{0pt}}m{#1}}
\newcolumntype{C}[1]{>{\centering\let\newline\\\arraybackslash\hspace{0pt}}m{#1}}
\newcolumntype{R}[1]{>{\raggedleft\let\newline\\\arraybackslash\hspace{0pt}}m{#1}}

\usepackage{booktabs}

\newcommand{\xhdr}[1]{\vspace{2mm}\noindent{{\bf #1.}}}

\newcommand{\R}{\mathbb{R}}

\begin{document}

\setcopyright{acmcopyright}

\doi{}

\isbn{}

\conferenceinfo{MLG '16}{August 14, 2016, San Francisco, CA, USA}


%

\title{Predicting risky behavior in social communities}

%
%
%
%
%

\numberofauthors{2} 
%
\author{
%
%
\alignauthor
Olivia Simpson\\
       \affaddr{University of California, San Diego}\\
       \email{osimpson@ucsd.edu}
\alignauthor
Julian McAuley\\
       \affaddr{University of California, San Diego}\\
       \email{jmcauley@cse.ucsd.edu}
}

\maketitle
\begin{abstract}
Predicting risk profiles of individuals in networks (e.g.~susceptibility to a particular disease, or likelihood of smoking)
is challenging for a variety of reasons. For one, `local' features (such as an individual's demographic information) may lack sufficient information to make informative predictions; this is especially problematic when predicting `risk,' as the relevant features may be precisely those that an individual is disinclined to reveal in a survey. Secondly, even if such features are available, they still may miss crucial information, as `risk' may be a function not just of an individual's features but also those of their friends and social communities. Here, we predict individual's risk profiles as a function of both their local features and those of their friends. Instead of modeling influence from the social network directly (which proved difficult as friendship links may be sparse and partially observed), we instead model influence by discovering social communities in the network that may be related to risky behavior. The result is a model that predicts risk as a function of local features, while making up for their deficiencies and accounting for social influence by uncovering community structure in the network.  We test our model by predicting risky behavior among adolescents from the \emph{Add health} data set, and hometowns among users in a Facebook ego net.  Compared to prediction by features alone, our model demonstrates better predictive accuracy when measured as a whole, and in particular when measured as a function of network ``richness.''
\end{abstract}

%
%


%
%

%
%



\section{Introduction}

The power of predictive analytics lies in the questions asked.  At a doctor's office, diagnoses, tests, or lab work are ordered based on the answers to questions given at a checkup.  In the case of adolescents, the doctor may also ask questions of the parents or caretakers.  The answers to these questions are the observed data upon which the doctor will perform some predictive task.  In many situations, this data provides enough information for a confident prediction.  However, there are cases for which there is an untapped wealth of information available in the social network that would greatly enhance the quality of the observed data. 

The motivation for this work is the question: what if learning about the individual is not enough?  In the case of human behavior, more can be learned by collecting information not only of that individual, but of their social connections as well.  In particular, risky behavior may be influenced by factors beyond the individual, and may be better detected, and thereby prevented, if knowledge of the social network were included in observed data.

One way to account for the deficiencies of data available from individuals is to model \emph{influence} that arises from their social connections (See, e.g.~\cite{christakis2008quitting,christakis2007spread}).
While a powerful approach, modeling direct influence may be challenging if the social network is sparse, or if `influence' arises not just due to individual social connections, but rather due to the collective effect of the groups of communities that individuals belong to.

In light of these concerns, we frame the problem of predicting risky behavior among adolescents in terms the communities they belong to. This means that we must build predictors that make use of local features, as well as features of the communities that individuals belong to, which must themselves be detected automatically.
Community detection is a fundamental problem in the study of social networks~\cite{fortunato2010community,malliaros2013clustering}. 
We cast the problem in terms of non-negative matrix factorization \cite{koren2009matrix}, which explains the observed adjacency structure in the network in terms of latent community vectors. Although we lack `ground-truth' to tell us whether these correspond to `real' social communities, we evaluate them in terms of their ability to improve the accuracy of the predictive tasks we consider.

We develop a model that incorporates both individual features and the social network for the purpose of label prediction.  We use this model to predict risky sexual behavior in adolescents and show that we are better able to predict in certain measures when incorporating the social network than when using features such as lifestyle, family, and general health (questions a doctor might ask) alone.  We test the model on a data set constructed from an in-depth survey of health, family, lifestyle, and relationships conducted on a nationally representative sample of adolescents from the United States to predict risky sexual behavior.  We additionally test our method on a Facebook ego net with profile metadata for each node.  Here, the feature data is sparse but there is a richly connected social network.

We compare our model to a baseline prediction using features exclusively.  We also use explicit features alongside the average of neighbor's features for comparison.  Our model performs better than the other two methods globally for the Facebook egonet.  For the Add health network we see improvement over the other models on a local scale.  In particular, when analyzing performance as a function of network information (measured by node degree), we see improvement in precision, recall, and accuracy at all scales of the network.



\subsection{Related work}

The most closely related branches of work to ours are 
(1) those that model social influence in networks
(2) those that model and summarize networks in terms of communities 
and
(3) those that solve predictive tasks defined on networks (or individuals within networks), especially with regard to health and risk-related applications.

\xhdr{Social influence} Modeling social influence in networks is a broad problem with many variants. 
One example is the spread of ideas, for example how tweets spread due to influence on Twitter (see e.g.~\cite{Bakshy}). More relevant to the type of model we develop is the idea of `social regularization,' where the network is used not just to model influence, but more critically to account for deficiencies (or missing data) available at the nodes themselves. This idea is especially popular in recommender systems, where data sparsity is a major issue (see e.g.~\cite{MaReg}).
More closely related to our work, we highlight a few examples that model the role of the network in relation to health data below.

\xhdr{Community detection}
Communities are groups of nodes with many internal connections, and relatively few connections to the rest of the network~\cite{ahn2010link,coscia2012demon,fortunato2010community}, and are well-studied in the sphere of network science.  Specifically, the problem of detecting such communities has been heavily explored and remains a fundamental problem, especially when considering social network data~\cite{malliaros2013clustering,mcauley2012learning,papadopoulos2012community}.

Some approaches to community detection are purely based on structure~\cite{fiedler1973algebraic,andersen2006local,chung2014computing}.  Generative models such as~\cite{yang2013community} incorporate node features in the generative process for predicting community membership as well.  Our method for community detection is based on matrix factorization.  The idea is not dissimilar from PCA approaches using the graph Laplacian~\cite{shen2012dimensionality} and so-called `signless' Laplacian~\cite{sarkar2011community}.

\xhdr{Health and the social network}
The Add health data sparked a number of studies on the relationship of the social network and human behavior and health.  For instance, in~\cite{strauss2003social} the authors look at social marginalization of overweight children, and in~\cite{ball2013friendship} social networks are used to learn social status rankings.  Each of these uses the Add health data set linking the social network to health.  

There are also a number of studies specifically on the role of social networks and risky sexual behavior (\cite{tyler2008social,morris1995social,ybarra2008risky}).  However, each of these papers focuses on an individual's interactions with social connections, and how this might affect engaging in risky sexual behavior.  In this paper, we examine how entire social communities and the individual's membership or relationship to these communities affects engaging in risky sexual behavior.

Longitudinal studies have also illuminated the effect the social network can have on human behavior.  In~\cite{christakis2007spread} and~\cite{bahr2009exploiting}, obesity is shown to spread through social ties.  Smokers are also shown to evolve in clusters~\cite{christakis2008quitting}, and in particular the authors show that clusters tend to quit smoking in concert.  These phenomena are also present among online social communities, as in~\cite{de2013predicting} postpartum depression is predicted with Twitter posts.  Our current work does not utilize the record of multiple phases of surveys over time, but it would be valuable to see how social communities affect the evolution of individual behaviors and we suggest this for further work.

\subsection{Contributions}

\begin{figure*}[t]
    \centering
    \begin{subfigure}[b]{0.4\textwidth}
        \centering \includegraphics[width=\textwidth]{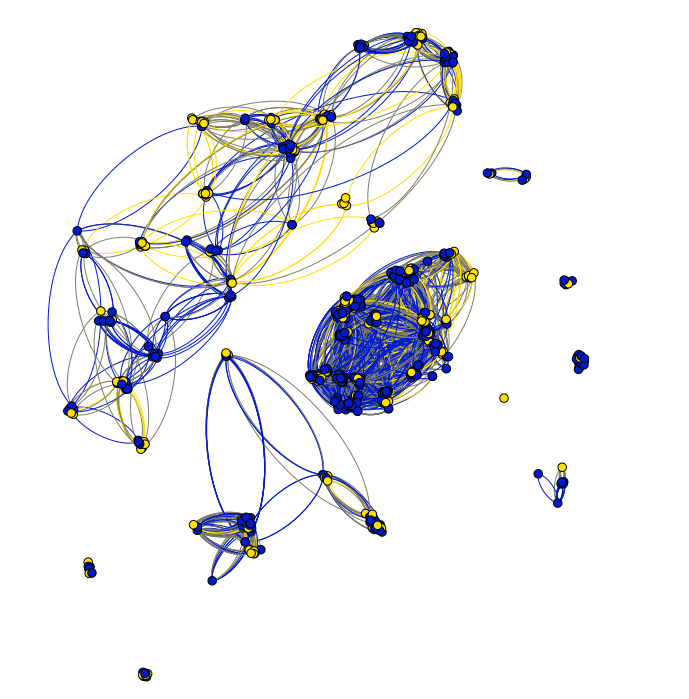}
        \caption{Actual labels}
        \label{fig:actual}
    \end{subfigure}
    \hspace{15mm}
    \begin{subfigure}[b]{0.4\textwidth}
        \centering \includegraphics[width=\textwidth]{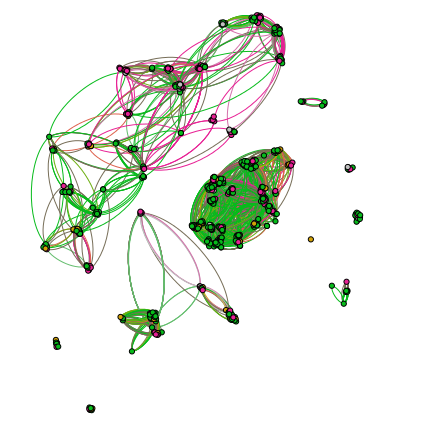}
        \caption{Predictions made with our model}
        \label{fig:lfm}        
        \end{subfigure}
    \caption{Visualization of the Add health social network.  Figure~\ref{fig:actual} depicts labels: yellow indicates risky sexual behavior, and blue indicates protected sex.  Figure~\ref{fig:lfm} depicts the predictions of our model: green indicates true positive, purple indicates true negative, yellow indicates false positive, and grey indicates false negative.}
    \label{fig:labels}
\end{figure*}

Our contribution builds on the ideas above from social influence, community detection, and predictive models in networks with health applications. Briefly, our main contributions are as follows:
\begin{itemize}
\item We build a model
to estimate missing features in attributed social networks. We apply this model to estimate the risky sexual behavior of adolescents on \emph{Add Health} data.  Figure~\ref{fig:labels} is a visualization of the Add health network with nodes colored by labels indicating engaging in risky sexual behavior (yellow) or not (blue).  We see that by comparing the actual labels to the predicted labels visually, our model demonstrates high accuracy.
\item Methodologically, our contribution is to use social network information to `fill in' missing information that cannot be predicted using node information alone. In contrast to existing work on social influence/social regularization, our predictor is based on (latent) community memberships rather than using the social links directly. This results in a robust predictor that accounts for the sparsity of the social networks observed in the Add Health dataset.
\item Our experiments reveal that by incorporating social community information, we have a higher predictive power by many measures of accuracy.  In particular, when decomposing error as a function of network connectivity as measured by node degree, we see that the predictions made by our model remain accurate at all scales on the Add health data as compared to predictions made by using features alone.
\end{itemize}

\section{The model}
We formally describe our model for predicting labels (engaging in risky behavior, for instance) based on individual features and the underlying social network.  Our notation is defined in Table \ref{table:notation}. Let $F$ be an $n \times f$ feature matrix with $n$ examples (individuals) and $f$ features.  By features we mean information pertaining to the individual.  We refer to the $i^{th}$ row of $F$ as the \emph{feature vector} $F_i$.  In general, the $i^{th}$ row of a matrix $M$, corresponding to the $i^{th}$ node, is denoted by $M_i$.  The social network of the examples is given by the $n \times n$ adjacency matrix $A$.  The social network should indicate the relationships of the $n$ individuals recorded in the feature matrix $F$.

The idea is to treat the network communities with respect to the individual as an additional feature.  That is, we aim to ``extend'' the feature vector by a vector encoding the individual's communities.  Thus, rather than using the full network topology, we use some low-dimensional representation of $A$ that encodes the communities of the network.  Let $C\in\R^{n\times k}$ with $k \ll n$ be a low-dimensional representation such that $g(C) \approx A$ by some transformation $g$.  The details of this reduction are given in Section~\ref{sec:dimred}, but for now let the $i^{th}$ row of $C$ be the \emph{communities vector} $C_i$.  Then, the full observed data for each individual consists of the feature vector and the communities vector.

Labels are predicted by incorporating the feature vector $F_i$ and the communities vector $C_i$ associated to example $i$ as $X_i := [C_i | F_i]$, where $[\cdot|\cdot]$ indicates column-wise concatenatation.  Namely, our matrix  of observed data incorporating the social network is $X = [C | F]$.  We perform logistic regression on $X$ for label prediction.

\begin{table}
\caption{Notation}\label{table:notation}
\begin{center}
\begin{tabular}{l l}
\toprule
$F$ & feature matrix\\
$A$ & adjacency matrix for social network\\
$X$ & full matrix of observed data\\
$n$ & number of examples\\
$f$ & feature dimension\\
$k$ & number of latent factors\\\bottomrule
\end{tabular}
\end{center}
\end{table}

\subsection{Network dimensionality reduction}\label{sec:dimred}
Our method for dimensionality reduction on $A$ is in the style of a latent factor model approach for learning the relationship of observed data by their latent factors.  Latent factor models are related to the singular value decomposition (SVD), but allow for optimization in spite of missing data. This approach also differs from the SVD due to the use of a link function (in this case a sigmoid), to account for the fact that the entries we are trying to estimate are in the range $[0,1]$. This is similar to community detection methods based on non-negative matrix factorization \cite{yang2012community} and so-called `one class' collaborative filtering \cite{rendle2009bpr}.


We model the basic probability that a link forms between example $i$ and example $j$ as $A_{ij} \approx U_iV_j^T$.  Certain biases may also be incorporated to enrich the model and to account for variance in the degree distribution of nodes in the network.  For instance, the likelihood that a link forms between example $X$ and example $Y$ may be slightly higher if example $X$ is generally popular, despite the individual relationship of $X$ and $Y$.  Or, if communication in general is very difficult within a particular social sphere, for example if coworkers are not encouraged or allowed to socialize in a particular workplace, the probability of links forming in general may be lower, despite individual relationships. We let $\beta$ be a vector encoding individual bias, and $\alpha$ a scalar encoding a global bias. In the first scenario, the row of $\beta$ corresponding to $X$ would encode this example's popularity, while in the second scenario $\alpha$ would encode a generally unsocial workplace.

Incorporating these biases, the full model for the probability of a link being added in the network is
\begin{equation}\label{costuv}
\Pr(A_{ij} = 1) = \sigma(\alpha + \beta_i + \beta_j + U_iV_j^T),
\end{equation}
where $\alpha$ encodes a global bias, and $\beta \in \R^{n}$ is a vector encoding individual example biases.

\subsubsection{Learning network parameters}
The goal is to find $U, V, \beta, \alpha$ which are good approximations to the binary matrix $A$, $A_{ij}\approx \sigma(\alpha + \beta_i + \beta_j + U_iV_j^T)$, as in (\ref{costuv}).  This is accomplished by arguments that minimize the cost function:
\begin{align}\label{eq:costlfm}
\begin{split}
\min_{U, V, \beta, \alpha} \sum_{i,j} & \frac{-A_{ij}}{2\omega_A} \log(\sigma(H_{ij})) -
\frac{(1-A_{ij})}{2\zeta_A} \log(1-\sigma(H_{ij}))\\
& + \gamma(||U_i||^2 + ||V_j||^2 + \beta_i^2 + \beta_j^2),\\
\end{split}
\end{align}
where
\begin{itemize}
\item $H_{ij} = \alpha + \beta_i + \beta_j + U_iV_j^T$,
\item $\omega_A := |\{A_{ij} ~|~ A_{ij} = 1\}|$, the number of edges
\item $\zeta_A := |\{A_{ij} ~|~ A_{ij} = 0\}|$, the number of non-edges, and
\item $\gamma$ is a regularization parameter.
\end{itemize}

Validation is used to determine the best value for $k$, tested in the range $(0,10)$, and regularization parameter $\gamma$.

\subsection{Using the model for prediction}

In this section we describe the specifics of performing a predictive task with the model.

We begin by reducing network dimensionality and learn relevant latent factors as described in Section~\ref{sec:dimred} by minimizing the objective cost function (\ref{eq:costlfm}) with respect to latent factor matrices $U$, $V$, individual bias vector $\beta$, and global bias constant $\alpha$.  As this optimization aims to approximate the adjacency matrix, $A$, the function is minimized over a training set consisting of half of the edges in the graph and two non-edges for every edge.  That is, the training set $C_{\mathit{train}}$ consists of $n_{\mathit{train}} := 1/2 \omega_A$ distinct random members of $\{(i,j) ~|~ A_{ij} = 1\}$ and $2n_{\mathit{train}}$ distinct random members of $\{(i,j) ~|~ A_{ij} = 0\}$.

Optimal values of $k$, the number of relevant latent factors, or the number of columns of $U$ and $V$, and regularization constant $\gamma$ are determined over a distinct validation set $C_{\mathit{val}}$ consisting of $n_{\mathit{val}} := 1/4 \omega_A$ edges and $1/2 \omega_A$ non-edges.  The best values of $k$ from $\{5, \ldots, 10\}$ and $\gamma$ from $\{0, 0.01, 0.04, 0.16, 0.64, 2.56, 10.24\}$ are those that optimize accuracy of the link predictions on the validation set.  Finally, the factors are tested on a withheld test set consisting of the remaining $n_{\mathit{test}} := 1/4 \omega_A$ edges and another $1/2 \omega_A$ non-edges.

After learning latent factors $U$, $V$, we construct the matrix $X = [U|V|F]$ which concatenates column-wise.  From this set of observed data we perform logistic regression to learn model parameters $\Theta$, and use threshold $t = 0.5$ to predict individual labels.

We divide the data into a training,
validation,
and test sets
in a 60/20/20\% ratio.
In each case we train the model on the training set using different
regularization parameters $\in \{0, 0.01, 0.02, 
\ldots
2.56, 5.12, 10.24\}$.  We choose the parameter that yields the highest binary classification rate on the validation set. 
We report all results on the test set.
 
\section{Data}\label{sec:data}
We test the model on two data sets.  The first, Add health, is an extensive study of family, health, and relationships among adolescents.  The second is the egonet of a Facebook user with accompanying metadata.

The main interest of this work is in predicting risky behavior among adolescents using the Add health data.  For the sake of comparison, and to assess the generalizability of our approach to other tasks, we also compare the performance of our model on the egonet, whose features are potentially less meaningful (school, location, workplace, etc.) but whose social network is comparitively richer (in particular, more complete) compared to the Add health dataset.

\subsection{Add health}\label{sec:addhealthdata}
The National Longitudinal Study of Adolescent to Adult Health (Add Health)~\cite{addhealth} is a longitudinal study of a nationally representative sample of adolescents in grades 7-12 in the United States.  The study collects data on physical and psychological well-being, as well as social and economic status.  The data also contains contextual information on the family, neighborhood, community, school, friendships, peer groups, and romantic relationships of the students.  According to the Add Health project summary, a major goal of the study is ``providing unique opportunities to study how social environments and behaviors in adolescence are linked to health and achievement outcomes in young adulthood.''~\cite{addhealth:site}

The Add Health study is conducted in waves of interview questions, the first of which was done in 1994-1995.  The first wave collected 2,820 answers of 20,745 students, for a data matrix of size $21K \times 3K$.

A crucial section of the study asks students to name friends of both genders---either one friend of each gender or 5 friends of each gender.  The observed data consists of this social network which contains edge labels indicating the type of relationship, as well as the answers to the interview questions.

\xhdr{Data pre-processing}
To ease computation and focus, we use the $k$-cores of the Add Health graph ($k = 9$).  A $k$-core of a graph is a maximal connected subgraph in which all vertices have degree at least $k$.  The collection of all $k$-cores will be referred to singularly as the $k$-core.  The $k$-cores contain a subset of the most active individuals in the group.

The $k$-core of the Add Health graph consists of 587 nodes and has an edge density of $0.012$, compared to the full graph of 107K nodes and an edge density of $7.2e-05$.  Properties of the $k$-core are given in 
Table~\ref{table:netstats}.  We note that while the original network is directed, the $k$-core is computed on the undirected version of the graph.  In particular, degrees of nodes are computed as the sum of in-degree and out-degree.

%

In our experiments, we aim to predict risky sexual behavior among adolescents.  Specifically, we predict the ``yes'' or ``no'' answer to the question given in the questionnaire as: \emph{``Did you or your partner use birth control the first time you had sexual intercourse?"}\footnote{In particular, a positive instance is an individual who uses birth control.  Thus, our model is actually predicting safe sex, but allows for detecting risky behavior as well.}  As not all students are sexually active, we limit the examples we use to individuals who provided a meaningful ``yes'' or ``no,'' and appropriately pruned the social network to include only these examples as well.  We then compute the $k$-core of the resulting network, and use only the remaining 587 examples for our predictive task.  The resulting feature matrix $F$ is of size $587 \times 3570$ and $A$ of size $587 \times 587$.

\subsection{Facebook egonet}\label{sec:fbdata}
Our second data set is the egonet of a Facebook user along with the corresponding profile data \cite{mcauley2012learning}.  The profile data contains information such as school, degree, workplace, location, and languages the user knows.  Details of the network data are given in Table~\ref{table:netstats}.

We perform a somewhat more benign predictive task on the Facebook egonet by predicting the hometown of the individuals.  Specifically, we review the hometowns of each individual and choose the hometown with the most positive examples as the predictive feature.

\begin{table}
\caption{Basic network statistics}
\label{table:netstats}
\centering
\begin{tabular}{l | C{2cm} C{2cm}}\toprule
                & Add health $9$-core & Facebook egonet\\\hline
number of nodes & 587                 & 238\\
number of edges & 4122                & 8420\\
density         & 0.01                & 0.15\\
transitivity    & 0.26                & 0.77\\
number of components & 11                  & 6\\
max degree      & 56                  & 184\\
average degree  & 14.04               & 70.8\\\bottomrule
\end{tabular}
\end{table}

\section{Experiments and analysis of results}
We compare the results of our method (performing logistic regression on the full matrix of observed data $X$) to two simple methods.  The first is a baseline method performing logistic regression on the feature matrix $F$ alone.  For the second, we use the basic features as well as the average of neighbors' features.  We call the matrix $F$ concatenated with columns giving the average of each feature $N$ and perform logistic regression on $N$.\footnote{All code is publicly available at \url{https://github.com/osimpson/community-lfm}}  We compare the predictive power of these methods in order to evaluate the advantage gained by including community features in the model.  We use global measures such as accuracy and root mean squared error to evaluate the overall output of the model.  We additionally investigate individual predictions as a function of node degree in order to see where the most improvement is gained in terms of network ``richness.''  We use a number of quantitative measures as described below.

Let $\hat{y}_i$ represent the prediction for $y_i$ and $\tilde{y}_i$ the confidence of the prediction.  Then we measure accuracy as the fraction of correct predictions:
\begin{equation}
\text{accuracy} = \frac{1}{n}|\{i ~|~ y_i = \hat{y}_i\}|.
\end{equation}
A higher accuracy means more correct predictions.  The \emph{root mean squared error} (RMSE) is defined as usual:
\begin{equation}
\text{RMSE} = \sqrt{\frac{1}{n}\sum_i (\tilde{y}_i - y_i)^2}.
\end{equation}
The \emph{precision}, or positive predictive value, of our set of predictions is the fraction of correct positive predictions to all positive predictions.  The \emph{recall}, or sensitivity, is the fraction of correct positive predictions to total positive instances.  The F1 score is the harmonic mean of precision and recall.  Each of these are a measure of accuracy, and a higher value indicates better predictions.

The \emph{balanced classification rate} (BCR) is an average of the rate of correct positive predictions and correct negative predictions.

Finally, we also use the log likelihood of the learned model $\Theta$ 
as a quantitative measure.

We compare the predictions of our method to the predictions of the feature matrix $F$ as well as the neighbor feature matrix $N$.  We use the same row-wise partitions in each case, the only difference being the columns of the data matrix depending on the method.

\xhdr{Add Health}
As discussed in Section~\ref{sec:addhealthdata}, we predict whether or not individuals will engage in protected sexual intercourse during their first encounter.  Specifically, we predict the ``yes'' or ``no'' answer to the question given in the questionnaire as: \emph{``Did you or your partner use birth control the first time you had sexual intercourse?"}

Each of the $U$ and $V$ latent factor matrices are of size $587 \times 9$ where the number of relevant latent factors, $k=9$, is determined by validation on a withheld set (see Section~\ref{sec:dimred}).  We then compare the binary predictive performance of our model, using the full data matrix $X = [U|V|F]$ to the binary predictive performance using just $F$.  

\begin{figure}[t]
\begin{center}
\includegraphics[width=1\linewidth]{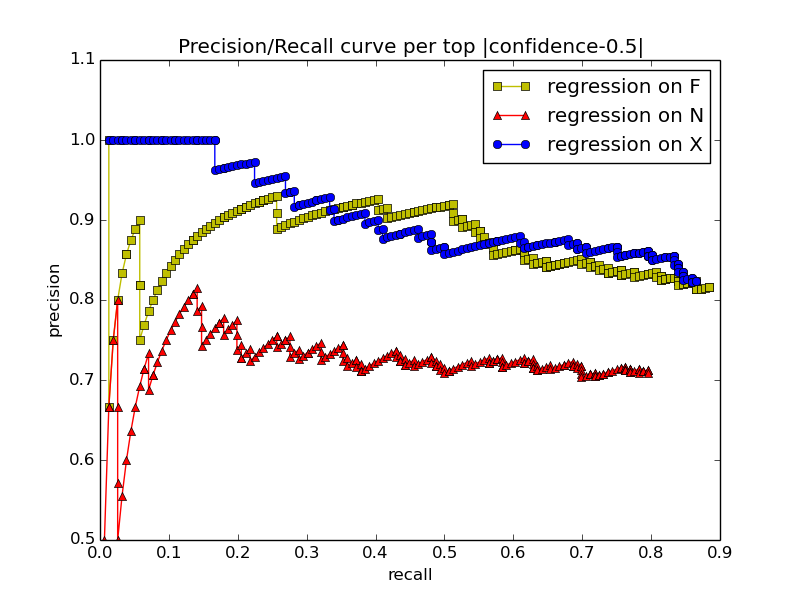}
\caption{Precision/recall curve for Add health data.  Area under the curve representing regression on $F$ is 0.7596.  The area under the curve representing regression on $N$ is 0.5712.  The area under the curve representing regression on $X$ is 0.7749.}
\label{fig:addhealth_recall_precision}
\end{center}
\end{figure}

\begin{figure}[t]
\centering
\includegraphics[width=1\linewidth]{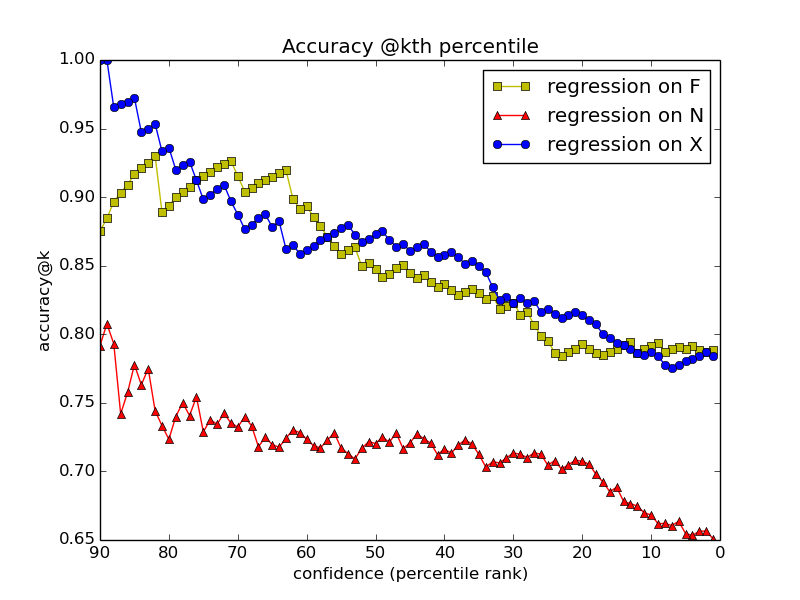}
\caption{Accuracy @kth percentile for Add health data}
\label{fig:addhealth_accuracy_atk}
\end{figure}

\begin{figure}[t]
\centering
\includegraphics[width=1\linewidth]{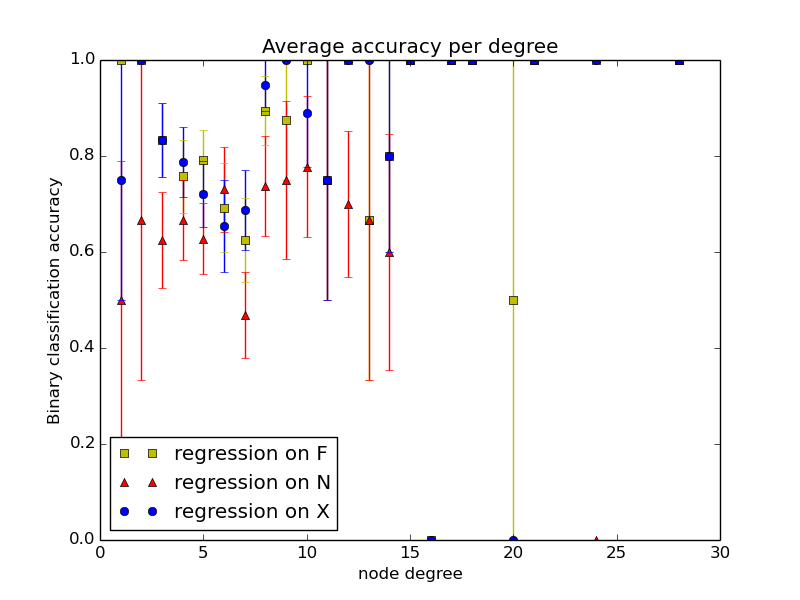}
\caption{Mean accuracy per degree with error bars representing one standard error for Add health data.}
\label{fig:addhealth_degree_accuracy}
\end{figure}

Table~\ref{table:addhealthresults} compares a number of performance measures for each of the models.  The first six, accuracy, log likelihood, precision, recall, F1 score, and BCR are all measures of accuracy, where higher values indicate better performance.  The next next is a measure of error, where smaller values indicate better performance.  We also look at the percent positive instances predicted.  Our model has better precision and BCR, though we see that, overall, the global measures are quite similar.  Small improvements over the baseline and using neighbor features at a global scale are further illuminated by decomposing the error.  We measure error at different scales to see where our model outperforms regression on $F$ and $N$.  In~\ref{fig:addhealth_recall_precision,fig:addhealth_accuracy_atk,fig:addhealth_degree_accuracy}, we plot these measures.

\begin{table}
\caption{Global performance measures for Add health}\label{table:addhealthresults}
\begin{center}
\begin{tabular}{l c c c}\toprule
& $F$ & $N$ & $X$\\\midrule
Accuracy           & 6.186e-1    & \bf{6.356e-1}     & 6.186e-1\\
Log likelihood     & 119.63      & \bf{141.71}     & 118.38\\
Precision          & 6.85e-1     & 6.91e-1         & \bf{7.02e-1}\\
Recall             & 8.0e-1      & \bf{8.2e-1}     & 7.47e-1\\
F1 score           & 7.368e-1    & \bf{7.514e-1}   & 7.239e-1\\
BCR                & 5.269e-1    & 5.396e-1        & \bf{5.529e-1}\\\hline
RMSE               & \bf{5.207e-1} & 5.369e-       & 5.282e-1\\\hline
\% pred. positive  & 0.78     & 0.8            & \bf{0.71}\\
\% actual positive     & 0.67     &                &
\\\bottomrule
\end{tabular}
\end{center}
Global performance measures of the predictions made by regression on $F$, the feature matrix, regression on $N$, the feature matrix including neighbor features, regression on $X$, the full observed data including the latent network community.  The first six measures are measures of accuracy, where a higher value indicates a better prediction.  For each measure, the best result is indicated in boldface.
\end{table}

In Figure~\ref{fig:addhealth_recall_precision}, we measure the recall and precision of the top $r$ predictions ranked by $|\tilde{y}_i - 0.5|$ for varying $r$, and measure the recall vs. precision of these top $r$ predictions.  For our model, the precision remains high and the area under the curve is greater than that for the baseline by 0.015.  We note that our model maintains perfect precision for recall up to 0.2 as more examples are included.  In this range, the precision of the baseline is between 0.75 and 0.9.  Further, where regression on $N$ resulted in better global measures (Table~\ref{table:addhealthresults}), here we see a marked improvement using our method.

Figure~\ref{fig:addhealth_accuracy_atk} measures accuracy as a function of confidence.  That is, among the predictions in the top $k^{th}$ confidence percentile, we measure the predictive accuracy.  The figure demonstrates that in general, with the exception of some of the third quartile, the accuracy of our model's predictions are better by a margin of up to 0.15 than that of the baseline.  Most notably, in the $75-90^{th}$ percentile, our model demonstrates much better accuracy.  Again, our model greatly outperforms regression on $N$.

To more closely investigate the quality of the prediction with respect to the quality of the social network, in Figure~\ref{fig:addhealth_degree_accuracy} we plot mean accuracy per network degree, with error bars indicating one standard standard error.  We see improvement at almost all degrees with our model, and meaningful improvement among common degrees, those $\leq 10$.

We also check other measurements of accuracy and error per node degree.  Specifically we measure the BCR, F1 score, and something we refer to as the \emph{perplexity} (see the section below).  In each of these cases, the results using regression on $X$ are not significantly different from using regression on $F$ or $N$, and we do not include those reports in our analysis.

It would be meaningful to perform these measurements on the data as a function of data richness.  One possible measure of data richness is the extent to which there is missing data in any row, for example.  We would expect that, at a point where the feature data is lacking, but the network data is rich, our would perform particularly well.  However, in the case of Add health the data is fairly consistent and a measurement along this axis was not revealing to the power of the model.  Again, we do not include these reports in our analysis.


\xhdr{Facebook egonet}
In this case, we predict whether or not individuals are from a common hometown (see Section~\ref{sec:fbdata}) based on Facebook profile data.

Each of the $U$ and $V$ latent factor matrices are of size $238 \times 10$.  As above, Table~\ref{table:fbresults} compares the models by a number of measurements.  Our model has significantly better precision and recall as compared to the baseline, as well as a better BCR.  The log likelihood is somewhat better, while the F1 score, and RMSE are somewhat worse.  Remarkably, although our model and the baseline predict with the same accuracy, our model predicts the correct number of positive examples, while the baseline severely underpredicts the number of positive examples.  Further, regression on $N$ results in no positive predictions (which may explain the high accuracy).  These global measures demonstrate that in cases where feature data is sparse, a rich network with communities can enhance predictive tasks.

\begin{table}
\caption{Global performance measures for Facebook egonet}\label{table:fbresults}
\begin{center}
\begin{tabular}{l c c c}\toprule
& $F$ & $N$ & $X$\\\midrule
Accuracy            & 8.7755e-1 & \bf{9.1837e-1} & 8.7755e-1\\
Log likelihood      & 22.69     & 23.16          & \bf{28.15}\\
Precision           & 0         & 0              & \bf{0.7}\\
Recall              & 0         & 0              & \bf{0.7}\\
F1 score            & \bf{7.368e-1} & 0          & 7.239\\
BCR                 & 4.78e-1    & 5.0e-1        & \bf{5.92e-1}\\\hline
RMSE                & 2.92e-1     & \bf{2.7e-1}  & 3.19e-1\\\hline
\% pred. positive  & 4.1e-2  & 0              & \bf{8.2e-2}\\
\% actual positive     & 8.2e-2  &                &
\\\bottomrule
\end{tabular}
\end{center}
Global performance measures of the predictions made by regression on $F$, the feature matrix, regression on $N$, the feature matrix including neighbor features, regression on $X$, the full observed data including the latent network community.  The first six measures are measures of accuracy, where a higher value indicates a better prediction.  For each measure, the best result is indicated in boldface.
\end{table}

\begin{figure}[t]
\centering
\includegraphics[width=1\linewidth]{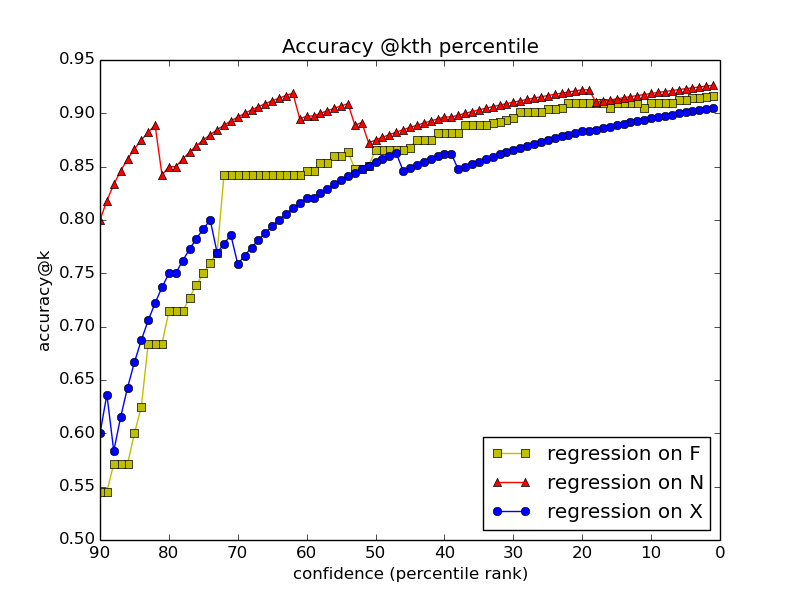}
\caption{Accuracy @kth percentile for the Facebook egonet}
\label{fig:ego_accuracy_atk}
\end{figure}

Again, we investigate performance at different scales.  Figure~\ref{fig:ego_accuracy_atk} measures accuracy @$k^{th}$ percentile.  Our model does consistently better than the baseline in the top quartile, but the baseline yields better accuracy elsewhere and regression on $N$ generally outperforms the other two.

\begin{figure}[t]
\centering
\includegraphics[width=1\linewidth]{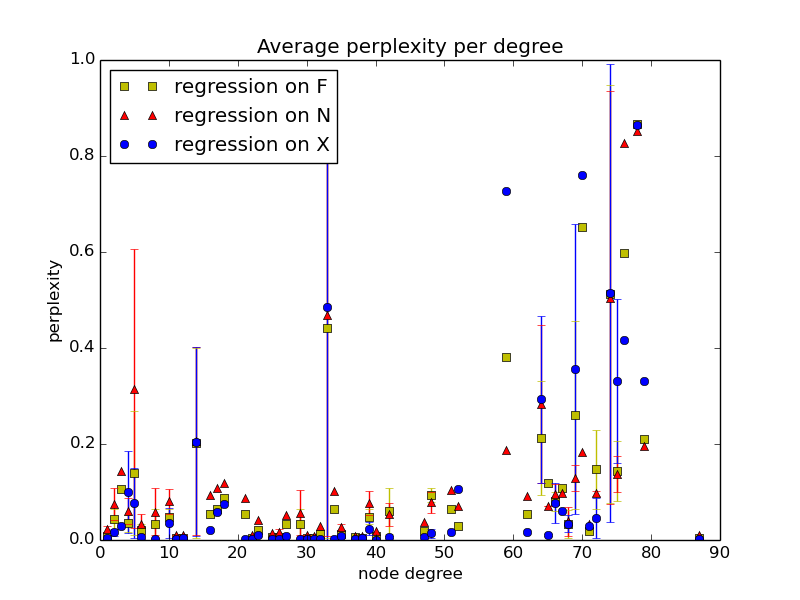}
\caption{Mean perplexity per degree for the Facebook egonet with error bars representing one standard error.}
\label{fig:ego_perplexity}
\end{figure}

In Figure~\ref{fig:ego_perplexity} we measure the average \emph{perplexity} per degree.  Again, if $\tilde{y}_i$ denotes the confidence of prediction $i$, and $y_i$ the actual label, then the perplexity of the prediction is $|\tilde{y}_i - y_i|$.  We see that our model demonstrates better (lower) average perplexity at almost every degree.

As with Add health, we additionally measure the accuracy, BCR, and F1 score as a function of node degree.  However, there is no meaningful difference here when using regression on $X$, $N$, or $F$, and we omit those reports.

\section{Conclusion}

This study illustrates two important ways the communities of a social network can help in a predictive task.  First, social communities inherently have an effect on individual behavior, and thereby health and well-being.  Incorporating social communities among the data for a predictive tasks adds valuable and measurably beneficial information that enhances the quality of the prediction.  We see the effects of social communities in our prediction of risky sexual behavior among adolescents with the Add Health data.  Second, if individual feature data is sparse or uninformative, social communities can boost the quality of the data used for a predictive task.  We see the improvement over feature data alone in predicting hometowns among individuals in a Facebook egonet.

The authors suggest a few avenues for future work.  In this work we use latent community structure as additional features, but this is done is something of a ``preprocessing'' step.  We believe that communities might be better detected if the node features were included in this process as well.  We believe a joint model which learns label prediction parameters and latent features simultaneously might enhance our model as it stands in this work.

Second, the Add health data set is a longitudinal data set.  It would be worthwhile to make use of this data to investigate how behaviors evolves over time with respect to social communities.

\section{Acknowledgments}
The authors would like to thank James Fowler for meaningful discussions in the preparation of this work.
%
\bibliographystyle{abbrv}
\bibliography{community}  
%
%
\appendix
Equation (\ref{eq:costlfm}) is minimized with gradient descent with the following
derivatives:
\small
\begin{align*}
\bullet& ~\frac{\partial}{\partial U_{ik}} =\\
& \sum_{j}\left( \frac{\sigma(H_{ij})}{2\zeta_A} -
A_{ij}\left( \frac{\sigma(H_{ij})}{2\zeta_A} +
\frac{1}{2\omega_A}\frac{1}{1+e^{H_{ij}}} \right) \right)V_{jk} + 2\gamma U_{ik}\\
\bullet& ~\frac{\partial}{\partial V_{jk}} =\\
& \sum_{i}\left( \frac{\sigma(H_{ij})}{2\zeta_A} -
A_{ij}\left( \frac{\sigma(H_{ij})}{2\zeta_A} + \frac{1}{2\omega_A}\frac{1}{1+e^{H_{ij}}} \right) \right)U_{ik} + 2\gamma V_{jk}\\
\bullet& ~\frac{\partial}{\partial \beta_i} =\\
& \sum_{j}\left( \left(
\frac{\sigma(H_{ij})}{2\zeta_A} - A_{ij}\left( \frac{\sigma(H_{ij})}{2\zeta_A}
+ \frac{1}{2\omega_A}\frac{1}{1+e^{H_{ij}}} \right) \right) +
2\gamma\beta_i\right)\\ 
&+ \sum_{j}\left( \left(
\frac{\sigma(H_{ji})}{2\zeta_A} -
A_{ji}\left( \frac{\sigma(H_{ji})}{2\zeta_A} +
\frac{1}{2\omega_A}\frac{1}{1+e^{H_{ji}}} \right) \right) +2\gamma\beta_i\right)\\
\bullet& ~\frac{\partial}{\partial \alpha} =\\
&  \sum_{ij} \left(
\frac{\sigma(H_{ij})}{2\zeta_A} -
A_{ij}\left( \frac{\sigma(H_{ij})}{2\zeta_A} +
\frac{1}{2\omega_A}\frac{1}{1+e^{H_{ij}}} \right) \right).
\end{align*}
\normalsize


\end{document}